\documentclass[10pt,twocolumn]{article}
\usepackage[letterpaper,left=0.75in,right=0.75in,top=1in,bottom=1in,columnsep=0.33in]{geometry} 
\usepackage[T1]{fontenc}
\usepackage[utf8]{inputenc}
\usepackage{newtxtext}   
\usepackage{newtxmath}   
\usepackage{microtype} 
\usepackage[htt]{hyphenat}   
\usepackage{booktabs}
\usepackage{graphicx}   
\usepackage{placeins}   
\usepackage{xcolor}
\usepackage[numbers,sort&compress]{natbib}
\usepackage[hidelinks]{hyperref}

\newcommand{\floornum}{\textbf{0.0}}
\title{\vspace{-2em}Trusted Floors Under Untrusted Learners:\\
A Runtime Assured-SLO Guard for ML Serving}
\newif\ifanon\anonfalse
\ifanon\author{}\else\author{Hsiu-Chi Tsai\\National Yang Ming Chiao Tung University\\\texttt{hctsai1006@cs.nctu.edu.tw}}\fi
\date{NSDI~'27 \emph{Frontiers} Track}

\begin{document}
\maketitle

\begin{abstract}
Modern ML serving increasingly lets \emph{learned, unverified} components (routers, latency-SLO admitters,
admit ladders) decide a tenant's quality of service; when one is wrong, the assured SLO can silently break, and
the Kubernetes layers beneath (Kueue, DRA, the Gateway-API Inference Extension, GAIE) add cross-layer
surprises. Rather than trust the learner to be right, we \textbf{bound the damage a wrong one can do}: a small
\emph{trusted} guard wraps the untrusted learner (\emph{learned proposes, the guard disposes}). A tenant's
assured-SLO obligation splits into two parts with different epistemics. Its \emph{safety projection}, a
per-class, per-window assured floor (with an optional drop rule, doom-sound only under an assumed service lower envelope), is a controllable obligation a guard \emph{enforces at runtime}, holding it
\emph{regardless of a learned admitter that is arbitrarily wrong within a bounded proposal interface} (it
proposes admission, class/objective selection, and priority hints, and actuates none of them against the assured floor). These guarantees are \emph{conditional}, scoped by
strength: the admission floor \emph{by construction} (modulo a shared window origin), the service
floor only \emph{conditionally}, under \S\ref{sec:guard}'s capacity, service-envelope, backlog, window-origin and scheduling preconditions. Its \emph{aggregate} obligation (the \emph{population} tail-latency
percentile) has no per-request enforcement point, so we treat it as a statistical residual and \emph{screen}
it. We build the guard (a Simplex-style assured-floor gate plus assured-first priority dispatch) and show on
real $2\times$V100 that it holds assured-class miss \floornum{} (admitted-basis) across \emph{every tested}
miscalibration of a learned admitter that, unguarded, misses $0.86$--$0.94$; against a live deployment of the
GAIE Flow Control, an injected mapping fault (emulating an untrusted mapper) flips the same assured requests from miss $0.0$ to
$1.0$ (a mechanism-level trust-boundary test, not a performance head-to-head) while our guard reserves by the
\emph{true} class. We also characterize \emph{when} a cheap static \emph{timing} screen can and cannot be trusted. As a
\emph{Frontiers} submission we evaluate the stance on commodity $2\times$V100 and a serving simulator, scoping
datacenter scale, real-model Flow Control, and a closed worst-case theorem as the agenda.
\end{abstract}

\section{Composition without assurance}\label{sec:compose}
A tenant's quality of service is no longer set in one place. Modern ML-serving stacks increasingly let
\emph{learned, unverified} components (smart routers, latency-SLO admitters, admit ladders) decide it at
runtime. That QoS is itself composed across four \emph{independently-configured} layers: cluster quota and priority
(Kueue's \texttt{ClusterQueue}/\texttt{WorkloadPriorityClass}), GPU device claims (DRA, \texttt{resource.k8s.io}),
inference-gateway priority and Flow-Control admission (GAIE's \texttt{InferenceObjective}), and
runtime scheduler knobs. When one component is wrong, the tenant's assured floor can silently collapse. Three
facts make this dangerous. \textbf{(i)} Cross-layer surprises are real: we catalogue
eight hazards where the \emph{composed} semantics surprise the operator (e.g.\ a DRA alias
double-counts a device; a Kueue \texttt{quotaCheckStrategy} flag alone flips admission of the
\emph{same} workload; App.~\ref{app:hazards}). \textbf{(ii)} Existing
tools miss them: \texttt{kubectl -{}-dry-run} and single-object OPA policies (scoped to the object under review, without synchronized cross-object inventory) validate one object
at a time and cannot compose these layers. \textbf{(iii)} The new mechanisms are learned and lack certified error bounds: GAIE's latency-SLO admitter, learned routers, and admit ladders provide empirical, not contractual, SLO evidence. In our tested sheddable-only policy, FIFO fails to protect the assured class under load \emph{even when the predictor is well-calibrated}.

The field's response has been to \emph{predict better}. We take a deliberately conservative stance. This
paper's central idea, an architectural principle: \emph{learned infrastructure controllers must not directly own contractual tenant entitlements}. Their proposals pass through a separately trusted entitlement boundary, and the deterministic contractual obligations are specified independently of statistical learner quality. Concretely, \textbf{bound the damage a wrong learned component can do, rather than trust it to be
right}. A small \emph{trusted} guard wraps the untrusted learner (\emph{learned proposes, the guard
disposes}) and holds each tenant's floor by its \emph{true} contract, however the learner errs; and we are
precise about which part of an SLO this stance even makes sense for.

\paragraph{Two obligations, two epistemics.}
The idea rests on a split: part of an assured SLO is a \emph{safety} property a runtime guard can \emph{enforce}
request-by-request, and part is a \emph{statistical} property it can at best \emph{screen}. Formally, model
executions as timed event traces; let $\hat{s}(R)$ be a service estimate the
system carries for request $R$---written $L(R)\!\le\!s(R)$ when it is a \emph{lower} bound and
$U(R)\!\ge\!s(R)$ when it is an \emph{upper} one, which are not interchangeable (below)---and fix a \emph{tumbling} window schedule $I_k\!=\![t_0\!+\!kW,\ t_0\!+\!(k{+}1)W)$ ($k\!=\!0,1,2,\ldots$) of length $W$ from a shared epoch $t_0$, and a per-window assured floor of
$B$ \emph{request credits} (in this prototype one request consumes one credit; token-weighted reservation, $\sum_{R} w(R)\!\le\!B$, is future work, though \S\ref{sec:guard} shows a \emph{count} floor already fails under variable sizes). A tenant's assured SLO bundles:
\emph{(a)} a \emph{safety projection}---\textbf{DR}, a \emph{doom-sound drop rule}: ``no request is
\emph{dropped} unless it is doomed'', sound \emph{given} a service lower bound $L\!\le\!s$ (our gate: drop $\Leftrightarrow t+L(R)>D(R)$); \textbf{AF}, ``in
every \emph{contract-aligned tumbling window} $I_k$ the assured class is admitted its first $\min(B,\,\text{offered in }I_k)$ requests from its
reserved credits, regardless of the learner'' (contract and guard share the epoch $t_0$; a mismatched epoch is a \emph{different} entitlement contract, quantified by \S\ref{sec:guard}'s phase sweep); and \textbf{SF}, ``every \emph{admitted} floor-entitled assured request
completes by its deadline'' (a \emph{per-request} guarantee, given the service envelope and capacity, \emph{not} a per-window completion \emph{count}, which fails under light demand). Each has a finite, irremediable bad prefix, so each is a
\emph{safety} property~\cite{alpern1985liveness,henzinger1992sooner}; and
\emph{(b)} an \emph{aggregate} obligation---\textbf{P95}, ``the $95$th-percentile
assured sojourn time \emph{over the arrival/service ensemble} $\le D$'', a \emph{distributional}
predicate over the trace ensemble (a \emph{probabilistic} hyperproperty~\cite{clarkson2010hyperproperties,wang2019statisticalhyper}, not a plain trace-set one), with no finite
bad prefix.

These demand different treatment. The safety projection is
\emph{runtime-enforceable}: not by Schneider truncation (which only
halts)~\cite{schneider2000enforceable} but as an edit/suppression
automaton~\cite{ligatti2005edit}: a serving guard \emph{suppresses} requests and
\emph{reorders} dispatch live, with \emph{controllable} admission in the sense of Basin et al.~\cite{basin2013enforceable}. The admission floor AF
is enforceable \emph{unconditionally against the learner} (modulo window-origin agreement, \S\ref{sec:guard}); the service floor SF holds only \emph{conditionally}: under capacity ($\ge\!B$ credits per window), a valid service envelope, bounded assured backlog, a shared window origin, and the deadline/scheduling preconditions of \S\ref{sec:guard}; DR is enforceable modulo the \emph{direction} of its estimate, and the
two directions are not interchangeable. With a lower bound $L$, dropping \emph{only} when $t+L\!>\!D$ is
\emph{doom-sound}: $t+s\!\ge\!t+L\!>\!D$, so \emph{under that hypothesis} every dropped request is infeasible and the guard never
sheds work it could have served. The converse property (\emph{no doomed request is dispatched}) instead needs
an \emph{upper} bound: dispatching \emph{only} when $t+U\!\le\!D$ gives $t+s\!\le\!t+U\!\le\!D$. What matters
is the \emph{direction} of the implication, not the test: ``drop \emph{only} when $t+L\!>\!D$'' (a drop
witnesses infeasibility) is doom-sound, whereas ``$t+L\!>\!D$ \emph{implies} drop'' is not: a gate that
dropped \emph{everything} would satisfy the latter and still shed feasible work. Under $L$ alone a request with $t+L\!\le\!D\!<\!t+s$ is doomed yet
undetected, and runs. Enforcing both would need a three-state gate (drop\,/\,dispatch\,/\,\emph{uncertain}); we
build the doom-sound direction and claim only it. The aggregate P95, by contrast, is neither per-trace
enforceable nor soundly provable by a deterministic first-moment model as
$\rho\!\to\!1$: the \emph{mean} sojourn diverges like $(1-\rho)^{-1}$ with
service-time variance (the Pollaczek--Khinchine mean~\cite{kleinrock1975queueing}), and under heavy-traffic scaling~\cite{kingman1961} so do its percentiles
(motivation, not a literal model of the continuously batched backend, whose variance such a first-moment model omits). \textbf{Enforce the safety projection; screen the statistical
residual}: the guard that enforces the first (\S\ref{sec:guard}) and a characterization of
when a cheap static screen can (and cannot) be trusted for the second (\S\ref{sec:screen}).

\paragraph{Early evidence the stance is promising.}
On real $2\times$V100 hardware the guard holds the assured floor (miss ${\approx}0$, reps $10$) around an
\emph{arbitrarily-miscalibrated} learned admitter, where the unguarded stack misses $0.86$--$0.94$. And against
a live deployment of the GAIE Flow Control (EPP v$1.5.0$, flow-control enabled), we
measure that this deployment dispatches strictly by the \texttt{InferenceObjective} priority; an \emph{injected} fault in the request-to-\texttt{InferenceObjective} mapping (emulating an untrusted mapper) flips the
\emph{same} assured requests from miss $0.0$ to $1.0$, whereas our guard reserves by the \emph{true} class, immune by
mechanism (a cross-reference to the V100 guard runs, not a same-stack head-to-head). The guard bounds the
damage a wrong learner can do through its proposal interface; the GAIE result is a separate mapping-sensitivity
finding (an injected fault propagating into strict-priority dispatch), not an observed failure of a deployed learned router.

\paragraph{Contributions: a Frontiers submission.}
\emph{(1)}~The \emph{stance}: a trusted floor that bounds an untrusted learner's damage, made precise by the
safety-vs-statistics epistemic split that says which part of an SLO it covers. \emph{(2)}~A guard realizing it,
with real-hardware evidence that it holds where an unguarded stack and a \emph{well-calibrated} FIFO admitter do not, plus a separate live-deployment fault-injection experiment where an erroneous class-to-\texttt{InferenceObjective} mapping propagates into GAIE Flow Control's strict-priority dispatch and causes the five tested assured requests to miss their deadlines. \emph{(3)}~A \emph{secondary}, honest characterization of when a cheap static screen can be
trusted for the residual, including a held-out result that it is a \emph{characterization}, not a certifier. We
evaluate on commodity $2\times$V100 and a serving simulator; datacenter scale, real-model Flow Control, and a
closed worst-case theorem are the aspects that are hard to evaluate, scoped as the agenda~(\S\ref{sec:open}).

\section{The guard: learned proposes, the trusted guard disposes}\label{sec:guard}
The guard's mechanism is \emph{classical} real-time/QoS scheduling: a per-class request-credit
reservation (in the spirit of CBS~\cite{abeni1998cbs}, but regulating request \emph{count}, not bandwidth) the opportunistic class can never draw from, $+$
assured-first non-preemptive fixed-priority dispatch, $+$ a firm-deadline drop
(deferred-red), wrapping the untrusted learned admitter/router. Protecting a reserved class
\emph{regardless of} a misbehaving lower-priority load is the defining guarantee of
reservation and \emph{mixed-criticality} scheduling~\cite{parekh1993gps,vestal2007mixedcrit};
we borrow the \emph{stance} of runtime assurance (Simplex: hold ``regardless of the outputs
of the advanced controller''~\cite{sha2001simplicity,blackboxsimplex,neuralsimplex}) but
\emph{not} its machinery: there is no decision module and no verified-fallback switch.
\textbf{Our claim is not the scheduler but its target}: the reserved object is the safety
projection \{DR, AF, SF\} of \S1, and the untrusted component is a
\emph{learned} serving admitter that is arbitrarily wrong \emph{within the bounded proposal interface
defined next}. So the assured floor is protected by
\emph{reservation} (AF) and \emph{priority} (SF), while deferred-red (DR), under an assumed service lower bound, keeps witnessed-doomed
requests off the server without shedding a feasible one (a near-inert limb here; see below).

\paragraph{What the ``true class'' is: the trust boundary.}
The reservation keys on the tenant's \emph{contract} class. That class, we assume, is fixed at the gateway from
the request's authenticated \emph{end-tenant} identity (a signed tenant claim or mTLS/workload identity; a Kubernetes ServiceAccount authenticates the gateway \emph{workload}, not by itself the end tenant of a shared inference request),
\emph{before} any learned component runs; the untrusted router/admitter sits \emph{downstream} of that binding
and only \emph{proposes}. Its output is $\pi_L(R)=(\text{admit\slash reject},\ \text{class\slash objective selection},\ \text{priority hint})$ (proposals only), with no actuation authority over the floor: floor-entitled assured is admitted from
\emph{reserved} credits whatever the learner proposes (in our runs the learner governs only the opportunistic
surplus, so its admit/reject cannot keep a floor-entitled request out, AF), and dispatch ranks on the
\emph{true} contract class, not the priority hint (SF). What the
learner cannot touch at all is the class key its floor is drawn from, the output cap, the batch width, placement,
and capacity. So the floor holds against a learner \emph{arbitrarily wrong within this bounded proposal
interface}---not against one that degrades capacity itself; a wrong \emph{class\slash objective-selection} proposal---a dispatch-class relabel, \emph{not} backend selection (which, like placement, would degrade capacity and is excluded above)---we measure separately
(\S\ref{sec:related}: the dispatch-class case reduces to admission$+$priority). This is the trust boundary the tested GAIE Flow Control deployment does not close (we do \emph{not} claim an inherent GAIE limitation): FC dispatches by the \texttt{InferenceObjective}
priority the routing layer \emph{selects} (via the request-to-objective mapping, not by authoring the priority value), so an \emph{injected} mapping fault (emulating an untrusted mapper) crosses straight into dispatch (our label-swap: the same
assured requests miss $0.0\!\to\!1.0$); our floor is keyed one layer earlier. The edge is therefore
\emph{conditional} on that binding: absent an authenticated class, no reservation---not even ours---can tell a
mislabel from the truth, so binding it end-to-end is a deployment assumption we state but do not close.

\begin{figure*}[t]
\centering
\includegraphics[width=\textwidth]{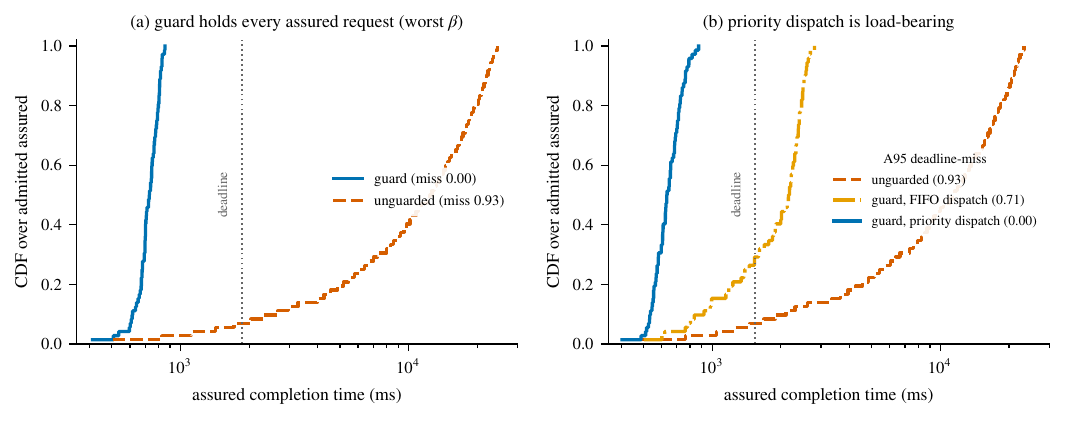}
\caption{\textbf{Reservation and priority, not the learner's calibration, hold the assured floor in the tested
operating region (AF structurally, SF under the stated service assumptions).} \textbf{(a)} CDF of assured completion, \emph{worst} miscalibration ($\beta\!=\!0$),
real $2\times$V100: every guard-admitted request completes $\le\!855$\,ms here ($\le\!924$ across the
$\beta$ sweep), far inside the $1856$\,ms deadline (miss \floornum{}); unguarded spreads past
$20$\,s (miss $0.93$). \textbf{(b)} Same CDF, a tighter fault-injection run (deadline $1544$\,ms),
\emph{byte-identical} admission: the unguarded arm again floods (both ${\approx}0.93$
coincidentally, each saturated under heavy overload), FIFO still misses $0.71$, only assured-first priority
pulls every request under it ($0.0$): priority, not headroom, is load-bearing. Log axis.}
\label{fig:guard}
\end{figure*}

\paragraph{It holds against an arbitrarily-wrong learned admitter.}
We drive a GAIE-style sheddable-only latency-SLO admitter as
the untrusted controller---a least-squares latency predictor swept over a miscalibration
$\beta$---on two Tesla V100s serving a
Qwen3.6-35B-A3B MoE model ($35$B total / $3$B active; AWQ $4$-bit, TP$2$, to fit $2\times32$\,GB V100s). The engine is the 1Cat-vLLM V100/SM$70$ fork (release v$0.0.3$, commit \texttt{72bb24e2d}, 2026-04-30; the installed package self-reports version $1.0.0$). Upstream vLLM lacks AWQ on sm\_$70$, so the fork adds TurboMind-derived $4$-bit kernels and a \texttt{FLASH\_ATTN\_V100} backend while leaving the scheduler stock, so its priority semantics are upstream's.\footnote{vLLM issue~\#40004, \url{https://github.com/vllm-project/vllm/issues/40004}.} Our GAIE-style sheddable-only reimplementation never sheds the class we designate assured, so the unguarded stack
\emph{overloads}: assured miss jumps $0.82\!\to\!0.93$ (single run; reps-$10$ below) then saturates, assured
queue-wait rising $1.5\!\to\!12$\,s. \textbf{The guard holds the assured floor at miss
\floornum{} for all four tested $\beta$} (Fig.~\ref{fig:guard}a plots the worst, $\beta\!=\!0$, run: worst \emph{observed}
assured completion $\approx\!855$\,ms in this run ($\le\!924$\,ms over all four $\beta$)\footnote{Service-number provenance (model, output-tokens, condition):
$2\times$V100 figures are Qwen3.6-35B-A3B at ${\sim}16$ tokens---$s_{\max}\!\approx\!660$\,ms (ceiling),
$386\!\to\!460$\,ms (service, light$\to$flood; $601$ max over the marquee), $855$--$924$\,ms (completion incl.\ queueing); ${\approx}214$\,ms
is $2\times$RTX~$6000$~Ada FP8; a batch-width probe measures $895\!\to\!1161$\,ms for the same Qwen at $64$ tokens.} $\ll\!1856$\,ms deadline, assured queue-wait $\sim\!130$\,ms, paired
opportunistic (A10, a $10\%$ assurance ratio) miss $\approx\!0.98$ (A95 $=$ the assured class, a $95\%$ assurance ratio). A \textbf{reps-$10$ companion $\beta$-sweep} ($D\!=\!1820$\,ms): the floor holds in \emph{every} one of the $40$ runs ($4\,\beta\!\times\!10$ reps; $0/2880$ assured requests missed, and \emph{none} rejected, so the offered-basis \emph{unmet fraction} equals the admitted-basis miss, both $0$, here), vs.\ unguarded $0.86\text{--}0.94$. The \emph{run} is the replication unit (as for the trace slices below): we observed \emph{no} failure across all $40$ runs, and each $\beta$ condition has $0/10$ with a two-sided $95\%$ per-condition upper bound of $0.31$ (we do not pool the $4$~$\beta$ into one Clopper-Pearson interval, they need not share a failure probability); the request-level Wilson $0.0053$ assumes independent requests and is optimistic under intra-run correlation. The empirical $0$ corroborates the \emph{structurally-enforced} AF and the \emph{conditional} SF (under the assumptions exercised here); it is not the basis of either. The protection is a property of the guard's
\emph{structure}: in a companion fault-injection run (guard admission at offered utilisation
$\hat{\rho}\!\gtrsim\!1$, i.e.\ offered load over capacity, FIFO vs.\ priority), dropping priority at
\emph{byte-identical} admission still misses $0.708$ (Fig.~\ref{fig:guard}b): \textbf{priority dispatch is
load-bearing, not headroom.} And even our \emph{well-calibrated} tested admitter does not protect assured under FIFO in this overloaded composition. Because the guard makes no static latency prediction, the
$\rho\!\to\!1$ knee that fools a static model (\S\ref{sec:screen}) has nothing to corrupt in
the guard: it is exposed to load, not to \emph{mis}-prediction. Two honest scopes: we bound a wrong \emph{admitter/router} (arbitrary admit/reject/classify under bounded output and a monitored service envelope), not a
capacity-degrading learner (placement, batching), since SF is capacity-conditional; and the
node is a single commodity $2\times$V100; like Chronos~\cite{chronos2026}, multi-node scale-out is future work.

\paragraph{Admitter-agnostic, honestly.} The floor does not depend on the learner's shed
policy. Driving the \emph{same} predictor ($132$-sample fit, $R^2\!=\!0.992$; the $\beta$
multiplier stresses \emph{miscalibration} full-range, so robustness isn't from the fit) under
a class-agnostic policy (vs.\ GAIE's sheddable-only), unguarded it fails at \emph{every tested}
$\beta$: by overload ($\beta\!=\!0$: admits $72$, miss $0.94$) or, a \emph{non}-GAIE case, by
rejecting assured ($\beta\!=\!1$: admits $29/72$, offered-basis unmet $0.94$); the guard holds miss \floornum{}
throughout. We do not oversell the rejection half: a reservation \emph{trivially} dominates an
under-admitting learner. The
load-bearing threat is \emph{overload}, held by the priority decomposition above
(Fig.~\ref{fig:guard}b), not by calibrating any learner.

\paragraph{The operating envelope (reps $10$, real $2\times$V100).}
A tighter sweep (reps $10$, run-level intervals; the companion $\beta$-sweep is above) hardens the
deadline/reservation axis. We report the \emph{measured}
operating fraction $\rho_{\mathrm{eff}}$ (admitted rate $\times$ measured service $/\,C$, the $C$ concurrent slots), distinct from the
theorem's worst-case $\rho_{\mathrm a}$ (on $s_{\max}$), and equal to the offered utilisation $\hat\rho$ below
saturation (admitted $=$ offered); since $\rho_{\mathrm{eff}}\!<\!\rho_{\mathrm a}$, A4 (defined below) is
\emph{sufficient}, not necessary, and the empirical break sits \emph{beyond} the guaranteed boundary. We directly measure the sustainable capacity $\mu_{\mathrm{sat}}\!=\!9.6$--$9.7$/s (saturating the $C\!=\!4$ server at the ${\sim}16$-token regime, $3$ reps); it falls inside the $C/\text{service}$ band $\rho_{\mathrm{eff}}$ normalises by ($8.7$--$10.4$/s), so $\rho_{\mathrm{eff}}$ and a $\mu_{\mathrm{sat}}$-based $\rho_{\mathrm{emp}}\!=\!\text{admitted}/\mu_{\mathrm{sat}}$ coincide: $\rho_{\mathrm{eff}}$ is a \emph{validated} load ratio \emph{for this} $C\!=\!4$, ${\sim}16$-token regime (the \S\ref{sec:screen} screen grids keep their configuration-specific normalized-load estimates), and the $\rho\!>\!1$ breaks are genuine overload past $\mu_{\mathrm{sat}}$, not a proxy artifact. Sweeping the assured
deadline at a sub-capacity reservation ($\rho_{\mathrm{eff}}\!\approx\!0.45$), the guard holds
miss $0.0$ \emph{down to a $600$\,ms deadline} ($0.003$ at $500$)---sub-capacity by design,
isolating the deadline axis---breaking only below the service (up to $0.49$ at $350$\,ms), a feasibility limit no admitter beats. On a \emph{separate} axis, raising
the reservation past capacity breaks the floor once the (unbounded) assured queue-growth
$\times$ horizon exceeds the deadline slack---clear at floor$=\!8$ ($\rho_{\mathrm{eff}}
\!=\!1.19$: miss $0.69$, ${\sim}190$\,ms/s), still absorbed at $\rho_{\mathrm{eff}}\!\approx\!
1.04$---mechanistically distinct from the deadline break. The
same sweep on a \emph{second} GPU ($2\times$RTX~$6000$~Ada, FP8, service ${\sim}214$\,ms)
holds the floor $0.0$ across \emph{every} tested deadline (to $350$\,ms): the floor
\emph{generalizes}. The base sweep held (inside both boundaries: $350\!>\!214$,
$\rho_{\mathrm{eff}}\!=\!1.07$); a follow-up past each reproduces the $\rho\!>\!1$ break
(floor$=\!22/26$: miss $0.52/0.78$), queueing physics, not a V100 artifact. \textbf{And the two stressors
compound}: at $\rho\!>\!1$ \emph{and} a tightening deadline the floor breaks
worse than either alone (V100 floor$=\!8$: $0.70\!\to\!0.99$ as $D$ tightens
$1200\!\to\!400$\,ms; the Ada node likewise), so the guarantee is a
\emph{region}: $\rho\!\lesssim\!1$ \emph{and} $D\!>$ service.
External validity: driving the \emph{live} $2\times$V100 vLLM end-to-end (each request is issued to the running engine, \emph{not} a simulator; arrivals fire \emph{open-loop} at the trace timestamps---the producer emits each request at its timestamp regardless of worker availability (the $4$ workers merely model the $4$ \texttt{max\_num\_seqs} slots), so backlog is server-side, confirmed below by the preserved fire-time CV and $\le\!2.3$\,ms lag) under a \emph{real} BurstGPT trace~\cite{burstgpt2025kdd}
(real bursty arrivals, interarrival CV~$2.94$ vs Poisson $1.0$, \emph{verified preserved at the guard} (realized fire-time CV~$2.94$, lag $\le\!2.3$\,ms, identical peak burst; \texttt{run\_p2\_mc4\_openloop.py})---the fully-real \emph{arrival} axis; output lengths capped at
$48$ tokens with the majority at the cap, input length not varied), the floor \emph{generalizes across two
models}, Qwen3.6-35B-A3B (MoE) and Mistral-7B (dense): the guard's \emph{admitted} floor all meet vs
admit-all's drowning (offered-basis \emph{unmet fraction} ${\approx}0.33$ vs ${\approx}0.9$, reps $3$); the $0.33$ residual is burst-excess
assured beyond the fixed floor $B$, a reservation-vs-burst bound, not a miss of admitted assured. Across $8$
non-overlapping slices of a separate bursty trace (reps $1$ per slice---the replication unit is the
trace slice) the guard's miss is burstiness-driven ($0.05$--$0.35$; the $8$ slices span interarrival CV
${\approx}0.85$--$2.75$, the burstiest missing most---though the rank correlation is not significant (tie-corrected Spearman $\rho\!\approx\!0.47$, $n\!=\!8$, $p\!=\!0.24$), so a range not a law; admit-all a flat ${\approx}0.9$), consistent
with the $0.33$ above (CV~$2.94$). Assumption A1's \emph{demote} (the theorem's regulator, which our experiments
approximate by the conservative \emph{reject}) recovers burst-excess at a loose deadline (deadline factor $2.5$):
the offered-basis \emph{unmet fraction} falls $0.35\!\to\!0.05$, so the reject bound is conservative. But it is a genuine
trade-off: at a tight deadline (factor $1.5$, $1230$\,ms) the demote's admitted excess contends for the $C$ slots, so
the reserved floor's met-fraction (met\,/\,floor-assured requests) is lower under demote ($0.69$ vs $0.77$, both outside the guaranteed region).
For deployment: default to the reject regulator (it protects the reserved floor at tight deadlines); switch to
demote only when the deadline is loose (deadline factor ${\gtrsim}2$), where it recovers burst-excess
without endangering the floor. (Scope: the run exercises
\emph{admission}; $\rho\!<\!1$ leaves priority and deferred-red inert.) Table~\ref{tab:basis} reports three
denominators side by side, the third being the one the contract actually names: \emph{floor-entitled
attainment}, the fraction of the requests entitled to the floor (the first $\min(B,\text{offered})$ assured per
window, AF) that are admitted \emph{and} on time. Reconstructed from the per-request raw
(\texttt{floor\_entitled.py}), it settles what an offered-basis number leaves open: \emph{w.r.t.\ the guard's own $W\!=\!1$\,s window} the guard rejects
\emph{zero} entitled requests in all three settings (every one of the $24$/$15$ BurstGPT rejections is
non-entitled burst-excess) and attains $1.00$, where admit-all attains $0.08$--$0.13$. Because entitlement is defined on tumbling windows it depends on the origin: a phase sweep (\texttt{floor\_entitled\_phase.py}, origin offset $\phi\!\in\![0,W)$) finds the marquee robust---attain $1.00$ at \emph{every} $\phi$, the guard rejecting nothing there---but the low-floor BurstGPT settings ($B\!=\!3$ Qwen, $1$ Mistral) alignment-sensitive: a contract window offset from the guard's would reclassify up to $18$/$9$ of those rejections as entitled (worst-case attain $0.67$/$0.73$). The floor-entitled guarantee is thus conditional on the guard's admission window sharing the contract's window origin.

\begin{table}[t]\centering\footnotesize
\setlength{\tabcolsep}{3pt}
\begin{tabular}{@{}llrrrccc@{}}
\toprule
setting & arm & off. & adm. & rej. & \multicolumn{3}{c@{}}{assured} \\
\cmidrule(l){6-8}
 & & & & & miss & unmet & attain. \\
\midrule
Learned-admitter & guard   & $2880$ & $2880$ & $0\%$  & $\mathbf{0.00}$ & $0.00$ & $\mathbf{1.00}$ \\
(marquee)        & admit-all & $2880$ & $2880$ & $0\%$  & $0.92$ & $0.92$ & $0.08$ \\
BurstGPT Qwen35B & guard   & $72$ & $48$ & $33\%$ & $\mathbf{0.00}$ & $0.33$ & $\mathbf{1.00}$ \\
                 & admit-all & $72$ & $72$ & $0\%$  & $0.92$ & $0.92$ & $0.13$ \\
BurstGPT Mistral & guard   & $45$ & $30$ & $33\%$ & $\mathbf{0.00}$ & $0.33$ & $\mathbf{1.00}$ \\
                 & admit-all & $45$ & $45$ & $0\%$  & $0.87$ & $0.87$ & $0.10$ \\
\bottomrule
\end{tabular}
\caption{\textbf{Three denominators for the assured class} (per-request raw; \texttt{basis\_breakdown.py}, \texttt{floor\_entitled.py}; marquee pools $4\,\beta\!\times\!10$ reps, unguarded per-$\beta$ $0.86$--$0.94$; BurstGPT pools reps~$3$). ``miss'' $=$ admitted-but-late\,/\,admitted; ``unmet'' $=$ (rejected $+$ admitted-but-late)\,/\,offered: a rejected request never runs, so it is not a latency \emph{miss}, but it is still \emph{unmet}; ``attain.'' $=$ floor-entitled attainment, (admitted \emph{and} on time)\,/\,floor-entitled; the BurstGPT $1.00$ is at the guard's window origin, alignment-sensitive (worst-case $0.67$/$0.73$ over window phase, \S\ref{sec:guard}), the marquee phase-robust. Admit-all shows why the third is needed: it satisfies the admission floor \emph{trivially} (it admits everything, AF $1.00$) and still attains only $0.08$--$0.13$. Attainment does not complement the miss where the denominators differ: on BurstGPT the entitled subset is the earliest arrival(s) of each window, which meet more often than the class as a whole ($0.13$ vs $1\!-\!0.92$). The miscalibrated \emph{learned} admitter (which instead \emph{rejects} assured at high $\beta$) is the separate \S\ref{sec:guard} run.}
\label{tab:basis}
\end{table}

\paragraph{Versus vLLM's own priority scheduler.}
A practitioner's first alternative to an admission guard is vLLM's server-side priority
(\texttt{--scheduling-policy priority}, admit-all, per-request priority; reps~$3$, $2\times$V100). It
\emph{also} holds the assured floor at the calibrated operating point ($D\!=\!1284$\,ms; miss $0.0$), but in our slot-bound (\texttt{max\_num\_seqs}$=\!4$) regime that priority does \emph{not} preempt a
running request (vLLM preempts only under KV/resource pressure; issue \#$40004$), so nearer the service time the
priority-jumped assured still wait for a running slot and slip (miss $0.15/0.26/0.46$ as $D$ tightens
$1100\!\to\!900$\,ms; per-rep consistent) where the admission floor holds $0.0$. Admit-all also \emph{self-congests} (unbounded server backlog; lower on-time opportunistic goodput under a hard opportunistic deadline).
The guard's edge over server priority (or GAIE Flow Control's priority$+$saturation detector) is
thus the per-class \emph{reservation}, not a dispatch rule or an aggregate shed.
A static rate limit (another native baseline) splits the same way (reps $8$): a \emph{per-class} cap
matches the guard (assured miss $0.0$), a \emph{flat} cap lets the opportunistic flood crowd assured out at
admission (miss $0.75$), so the load-bearing ingredient is the \emph{per-class reservation}, not rate-limiting per se. That count-based tie is size-scoped: with bimodal assured output lengths (a fraction large---here unservable at this deadline), the \emph{count} meter, blind to per-request cost, admits those large, which hold the $C$ slots non-preemptively and drag down the \emph{feasible} small assured ($39$--$51\%$ of them miss; admitted-basis SF miss $0.62$ real $2\times$V100 / $0.69$ deterministic-slot, reps~$3$). Those large \emph{are} floor-entitled under \S\ref{sec:compose}'s \emph{count} AF---a window offers exactly $B\!=\!3$ assured, so all three are among the first $\min(B,\text{offered})$, and the count meter admits them by the letter of AF; that is precisely why request-count is the wrong \emph{entitlement} meter here. A \emph{token}-metered analogue (the same GroupTB, admission cost $=$ the request's token weight $w(R)$---here the known synthetic output length, in practice an admission-time \emph{estimate}, not the realised length) instead sizes entitlement by cost: a request whose weight exceeds the floor budget is cost-\emph{excess} and is shed---which the count-based A1, seeing only $3\!\le\!B$, would \emph{not} do---so every feasible assured request the analogue admits meets its deadline (admitted-basis SF $0.0$), but only by rejecting those $41$ count-entitled large ($38\%$ of offered assured; offered-basis assured \emph{unmet fraction} $0.38$ in Table~\ref{tab:basis}'s sense---an AF trade under the count definition, not a clean win). And a coarse fixed token budget is itself only a \emph{proof-of-need}: the large are infeasible-alone, so it does not isolate cost-accounting from a plain feasibility gate (count $+$ feasibility check), and it would also shed feasible \emph{medium} requests---so a properly-sized token-weighted reservation is future work (\texttt{run\_p2\_tokenweight.py}).

\paragraph{Versus \emph{live} GAIE Flow Control.}
A strong $2026$ baseline is GAIE's Flow Control layer: a central priority queue with a saturation detector
and within-priority fairness keyed on a request-supplied fairness-id. We deployed one instance (Istio gateway, EPP v$1.5.0$ with \texttt{flowControl}
enabled, over vLLM-simulator backends; the EPP's FlowController confirmed running) and drove a saturating burst.
This deployment dispatches strictly by the \emph{declared} priority (high band miss ${\approx}0.31$ vs low band
${\approx}0.92$, reps $3$), so it protects whatever the routing layer \emph{labels} high priority. In a controlled mapping-fault injection at a fixed load, the \emph{same} five assured requests miss $0.0$ when labelled high vs
$1.0$ when the request-to-\texttt{InferenceObjective} mapping is swapped to the low band (both classes swapped; single run each): Flow Control obeys the \texttt{InferenceObjective.priority} it is handed---so a controlled fault we \emph{inject} into that mapping (emulating an untrusted mapper, not a deployed learned router) crosses straight into dispatch---and our tested configuration held no non-bypassable \emph{absolute} reservation tied to an \emph{authenticated} identity (that fairness-id is a mutable fair-share key, not a guaranteed floor, and even a correctly-mapped assured class has no absolute floor under enough
high-priority load). The guard's reservation, keyed to the \emph{true} contract class, holds regardless: the
fault-injection's opportunistic-first adversary is the direct analog (guard miss $0.0$; a no-priority variant
$0.708$). Backends are the official simulator, isolating the queueing mechanism; a head-to-head on one stack is
future work~(\S\ref{sec:open}). The deployment (Istio~$1.28$, EPP~v$1.5.0$ with \texttt{featureGates:[flowControl]}, three simulator backends) and raw are archived with the artifact.

\paragraph{A guard-floor bound.}
The fault injection gestures at a structural bound, what a \emph{structural} guarantee means
here: not a machine-checked $0.0$ (that number is empirical) but that \textbf{AF} is enforced \emph{structurally}; given A2--A4, a valid service envelope, a shared window origin, and the external backlog bound $Q_A\!\le\!B$, \textbf{SF} follows as a \emph{conditional} response-time implication (not by construction), while DR holds only under its assumed service lower bound. Assume \emph{(A1)} a reservation \emph{caps} admitted assured
at $B$ \emph{request credits} per \emph{tumbling} window $W$ (the prototype refreshes credits every $W\!=\!1$\,s; within a window this is a $\sigma\!=\!B$ cap, so at most $B$ assured are admitted) and reserves those $\le\!B$ a $\ge\!B$-credit capacity floor---the cap holds by construction, the floor is realized by (A2)--(A4) below given assured backlog (burst-excess handling is a policy choice---demote preserves it at opportunistic
priority, reject protects the reserved floor; the $\ge\!B$ floor holds under both) (AF); \emph{(A2)} non-preemptive assured-first priority
with arrival-order tie-break \emph{inside} the class (SF); \emph{(A3)} per-request service
(assured \emph{and} opportunistic) is bounded by $s_{\max}$ (per-class ceilings $s^{\mathrm{assured}}_{\max},s^{\mathrm{opp}}_{\max}\!\le\!s_{\max}$)---the batch-width cap
$C\!=\!\texttt{max\_num\_seqs}$ and a bounded output make this a \emph{plausible} ceiling,
though continuous batching still couples service to load \emph{within} it, which
\S\ref{sec:open} shows strains the assumption; and \emph{(A4)} the reservation is stable,
$\rho_{\mathrm a}\!=\!B\,s_{\max}/(CW)\!\le\!1$. While $R$ is queued, assured-first refill
dispatches \emph{no} new opportunistic, so $R$ waits only for one of the $C$
busy slots to free, an in-progress opportunistic job ($\le\!s^{\mathrm{opp}}_{\max}\!\le\!s_{\max}$);
then $R$ \emph{together with} the assured ahead of it---at most $B$ \emph{in total}, not $B$ besides $R$: the
$\sigma\!=\!B$ regulator caps the whole window's admitted assured cohort at $B$ credits and $R$ is one of them,
which is why the count below is $\lceil B/C\rceil$ and not $\lceil (B\!+\!1)/C\rceil$; this needs the assured
backlog not to accumulate across windows ($Q_A\!\le\!B$, a precondition made explicit below)---are served $C$-at-a-time by priority---the
\texttt{max\_num\_seqs}$=C$ batch serves $C$ logical sequence slots in parallel (under A3's service-envelope assumption), so serialization onto one slot is
\emph{not} assumed---in $\lceil B/C\rceil$ waves. Worst-case assured sojourn is thus
\[
  T_{\mathrm{assured}} \;\le\; s_{\max}\,\bigl(1+\lceil B/C\rceil\bigr)
\]
---the discrete makespan (an earlier $B\,s^{\mathrm{assured}}_{\max}$ \emph{sum} was loose;
a \emph{fluid} network-calculus form was \emph{unsound}, crediting an indivisible request only
$B/C$ of a slot). The floor
\textbf{meets this bound if (A4)~$\rho_{\mathrm a}\!=\!B\,s_{\max}/(CW)\!\le\!1$
(stability), $D\!\ge\!s_{\max}(1+\lceil B/C\rceil)$ (deadline), \emph{and} the assured backlog stays bounded ($Q_A\!\le\!B$)}. The $\sigma\!=\!B$ regulator bounds \emph{within}-window arrivals to $B$; that $Q_A$ does not accumulate \emph{across} windows is a backlog lemma we take as a \emph{precondition}, not a closed result (a counterexample would be a burst straddling a window boundary at $\rho_{\mathrm a}\!=\!1$); empirically it holds here---the assured backlog never exceeds $Q_A\!=\!3$ over the $40$ marquee runs ($B\!=\!6$) and $Q_A\!=\!1$ at the floor$=\!4\!=\!C$ corner cell below ($B\!=\!4$), each well under its floor (\texttt{envelope\_check.py}). \emph{In cell terms} at our operating point ($C\!=\!4$, $s_{\max}\!\approx\!660$\,ms---the
marquee's bounded ${\sim}16$-token-output ceiling): only the
loose-deadline floor$=\!4\!=\!C$ cells ($\lceil B/C\rceil\!=\!1$; $D\!\ge\!1320$\,ms, realized by the operating-envelope's floor$=\!4$ deadline-sweep cell at $D\!=\!1856$, guard miss \floornum{}, $\rho_{\mathrm{eff}}\!=\!0.45$) satisfy this bound \emph{conditional} on A1--A4, the backlog precondition $Q_A\!\le\!B$, and on treating the calibrated $s_{\max}$ as a valid service envelope---we do \emph{not} claim the measured $s_{\max}$ is a formally established WCET (though the $660$\,ms ceiling is never exceeded here---max observed assured service $601$\,ms over the $40$ marquee runs); away from the corner the guard still \emph{typically} holds $0.0$---the $B\!>\!C$ reservations to floor$=\!7$ ($\rho_{\mathrm{eff}}$ to $1.04$) and feasible deadlines to $D\!\approx\!600$, spread arrivals never sampling the co-arrival worst case---while the deepest overload (floor$=\!8$, $\rho_{\mathrm{eff}}\,1.19$, miss $0.69$) and sub-service deadlines break, the operating-region boundary the reps-$10$ V100 sweep above maps. \textbf{The floor rests on AF$+$SF, not DR.} We wire deferred-red as \S\ref{sec:compose}'s doom-sound rule (a
dispatch-time drop when $\mathrm{start}+L>D$), with $L$ the measured batch-$1$ \emph{median} service
${\approx}236$\,ms: the \emph{opposite} direction from the screen's $\hat s\!\ge\!s$ (\S\ref{sec:screen}), and
with the same epistemic status: a median over $12$ probes makes $L\!\le\!s$ \emph{assumed}, not established.
It holds where DR fires (the minimum \emph{batched} service we observe is $281$\,ms); ${\approx}0.4\%$ of
\emph{light-load} services dip to $234$\,ms, below $L$, but in cells where DR never fires. Under that
assumption DR sheds only the doomed, and even then does \emph{not} catch every doomed one; it is near-inert,
firing in $1$ of $15$ guard-DR cells under overload. So reservation$+$priority carry the floor \emph{independent of} $\hat s$; DR
is an \emph{optional}, conditional optimization for that tight-deadline/overload regime (safe only under a valid service lower envelope; no false drop observed where it fires; \S\ref{sec:screen}'s screen also turns unsound there), not a load-bearing safety property. \textbf{Where the tail is.} A3's load-\emph{independence} is strained on our vLLM backend:
continuous batching couples service to batch width, so opportunistic admission inflates the
\emph{assured} service too ($386\!\to\!460$\,ms under flood, same ${\sim}16$-token regime), yet this stays
\emph{below} $s_{\max}\!\approx\!660$\,ms, so the \emph{ceiling} the bound actually needs still holds. What
the coupling breaks is the clean ``learner enters through $s^{\mathrm{opp}}_{\max}$ alone''
\emph{separation}, not the conditional response-time bound. A true \emph{co-execution} cap (not a token
cap, which bounds tokens, not service) would restore the separation and tighten $s_{\max}$; that
batch-coupled tail (not the $924$\,ms worst-\emph{observed}) is the open problem
(\S\ref{sec:open}), and the CPS-Simplex lineage has no such service channel, so the transplant is not free.

\section{When can a cheap static screen be trusted?}\label{sec:screen}
The guard (\S\ref{sec:guard}) protects a \emph{live} request; the ecosystem will
nonetheless want to catch cross-layer misconfigurations \emph{before} they ship, with a
cheap static check. This section asks a subordinate question (\emph{where can such a
check be trusted}) and answers it empirically. We provide an offline screen: each layer modeled as a transfer function
pinned to the shipped component by trace-equivalence conformance, composed in a
discrete-event simulator that returns a counterexample trace or declines. It is
\emph{not} a prover and \emph{not} a sound abstract interpretation~\cite{cousot1977ai}: the $p50$ screen is \emph{optimistically} unsound (it false-accepts), becoming conservative only under a validated service percentile (below). We report it in the soundiness tradition~\cite{livshits2015soundiness}: an empirical screen with \emph{measured} false-accept regions. Two honest facts scope it. First, it is
\emph{two disjoint analyses}: a timing simulator (below) and a separate
quota-conservation arithmetic (the double-count/silent-skip hazards); the
timing screen catches no config hazards, and the quota screen is not
hardware-validated. Second, its timing verdict is fed a \emph{measured} service
estimate, so its accuracy figures are post-dictive. A deployable use needs a
pre-deployment $\hat{s}$ with the side-condition $\hat{s}\!\ge\!s$, obtainable from a short
batch-$1$-to-$C$ calibration probe (batch-$1$ ${\approx}236$\,ms; batch-$C$ ${\approx}590$\,ms
median under sustained full-batch load, $2\times$V100) run \emph{once before} the workload. Feeding a high batch-$C$ percentile
(padded to the ceiling $s_{\max}\!\approx\!660$\,ms) is our operational stand-in for the conservative
$\hat{s}\!\ge\!s$ the screen needs, at some completeness cost (over-estimating the \emph{typical} service). We
are explicit that this is an \emph{empirical envelope}, not a proven bound: a sample percentile bounds the
\emph{calibration sample}, not every future request, so $\hat{s}\!\ge\!s$ is assumed, not established, and the
held-out result below measures what a thin calibration costs. The residual is the
batch-coupled \emph{under-load} service (\S\ref{sec:guard}), which the probe \emph{covers over its calibration range} but does not \emph{predict} for future requests.

\begin{figure*}[t]
\centering
\includegraphics[width=\textwidth]{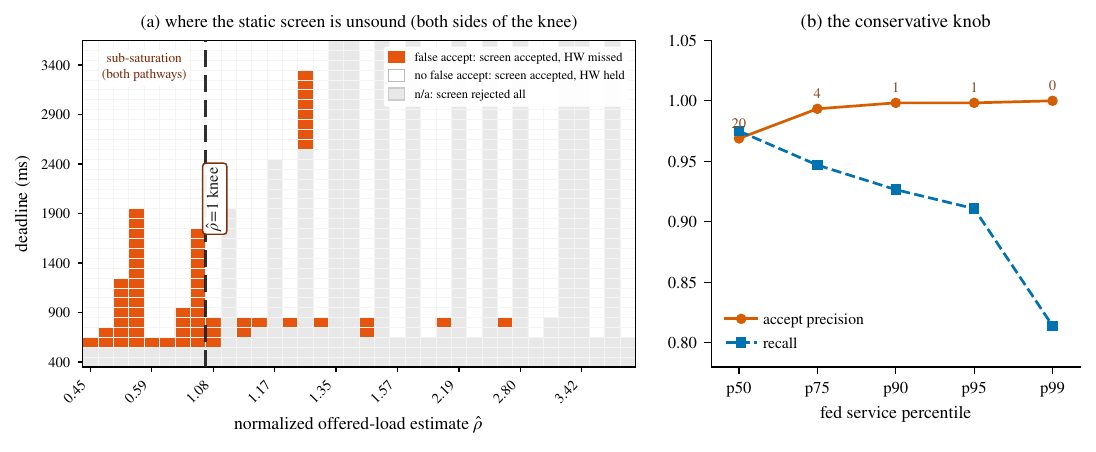}
\caption{\textbf{Where the cheap static screen is unsound: the hardware grids, aggregated.}
\textbf{(a)} Each $(\hat{\rho}\times\text{deadline})$ bin is \emph{orange} where the screen \emph{accepted}
a config the hardware then \emph{missed} (a false accept, so unsound \emph{there}), white where an accepted
config held, grey where it rejected all (n/a). The
$\hat\rho$ axis now spans both grids ($0.45$--$4.06$, reps-$10$ sub-saturation $+$ reps-$3$ overload, unmixed); false-accept cells sit on
\emph{both} sides of the $\hat{\rho}\!=\!1$ knee (dashed)---the overload knee and deep sub-saturation---and along
the tightest deadlines. \textbf{(b)} A conservative percentile
drives the $1320$-grid's observed false accepts to zero (precision $0.969\!\to\!1.000$, recall $0.975\!\to\!0.814$); near $\hat{\rho}\!\to\!1$ only the runtime guard holds.}
\label{fig:envelope}
\end{figure*}

\paragraph{The characterization.} Across $40$ configurations and $33$ deadlines
($1320$ cells, all in overload $\hat{\rho}\!\in\![1.08,4.06]$; Fig.~\ref{fig:envelope}), the timing screen's accept verdict tracks
the measured assured-SLO outcome at precision $0.969$ / recall $0.975$ (config-clustered $95\%$ CI $[0.94,0.99]$ / $[0.96,0.99]$, a bootstrap over the $40$ configs since the $33$ deadlines per config are correlated re-thresholds, the $20$ false accepts concentrate in $9$ of $40$ configs, up to $8$ in one; agreement $0.973$ is base-rate-inflated,
as the always-reject stub below shows; a service-only baseline scores $0.500$).
The $1320$ is not $1320$ hardware runs: $120$ are ($40$ configs $\times\,3$ reps), the $33$ deadlines
swept \emph{post-hoc} by re-thresholding \emph{measured} completion times (real, not simulated);
the verdict is conformance-pinned to the shipped admission/dispatch.
The conservative $p99$ service percentile is that empirical envelope made operational. But
the $p50$ accept verdict is \emph{near-worthless} on contested near-capacity cells
(accept-precision $0.167$; an always-reject stub wins there, $0.983$ vs $0.948$); $p99$ drives the
whole grid's \emph{observed} false accepts to zero (precision $1.000$ on this grid) only by over-flagging (recall $0.975\!\to\!0.814$): it
rejects the contested cells outright, vacuously so at near-zero coverage. The $p50$ verdict's $20$ over-optimistic errors are variance-driven ($8$
gated-FIFO at the $\hat{\rho}\!\approx\!1.18$ knee, $12$ priority at tight deadlines): \textbf{the finding is the boundary}: across this overload grid the accept verdict has no \emph{observed} false accept except there. Below saturation we now \emph{measure} it: the $p50$ screen is optimistically unsound there too: precision
$0.83$ at $\hat{\rho}\!<\!1$ (mean-aggregation; gated, opp$\in\!\{1,2,3\}\times$slots$\in\!\{4,6,8\}\times\{$FIFO,\,priority$\}$, $18$ configs
$\times\,33$ deadlines, reps 10)\footnote{Precision by regime$\times$aggregation, with accept counts: sub-saturation ($N\!=\!248$ accepts) $0.79/0.83/0.93$ and same-sweep overload ($N\!=\!189$) $0.89/0.94/0.99$ under max/mean/median aggregation. A cell's $\hat\rho$ is classified by the reps-$10$ median service; the $64\%$ coverage below is $168/264$ cells over the $8$ sub-saturation configs (of the $18$) $\times\,33$ deadlines; the $42$ false positives span \emph{all} $8$ configs.} vs $0.94$ in the same sweep's overload cells
(near the knee, $\hat\rho$ to $1.36$; not comparable to the deep-overload $1320$-cell grid's $0.969$), false positives down to $\hat{\rho}\!\approx\!0.45$ (stable across the
$10$ reps---$39/42$ miss in $\ge\!3$, none transient)---by two pathways
of the same service dispersion a first-moment model omits: the service tail entering the sojourn \emph{directly}
in deep sub-saturation where the queue is empty (the same below-service-time infeasibility \S\ref{sec:guard}'s
feasibility limit names), and the $(1-\rho)^{-1}$-amplified service variance of \S1 \emph{in the wait} nearer the
knee. Which dominates is sensitive to the service estimate, so we claim only that both contribute. A conservative
$p99$ service percentile closes them: fed $p99$, the sub-saturation false positives fall to zero (precision
$1.000$) while the screen still \emph{accepts} $64\%$ of cells, the same conservatism/coverage dial that closes the
overload accepts ($0.969\!\to\!1.000$; $p99$ the aggregation-robust choice)---but this is a \emph{post-hoc} patch: held-out on the sub-saturation reps ($\hat{s}$ from \emph{one} held-out rep $\approx\!10\%$, tested on the unseen $9$, rotated $10$-fold, run-level disjoint) the $p99$ closure weakens (precision $1.000\!\to\!0.90$ at $10\%$ calibration, recovering to $0.99$ near $50\%$---a calibration-\emph{data} requirement, not irreducible): a \emph{characterization}, not a pre-deployment certificate, at a ${\sim}36\%$ coverage cost (\S\ref{sec:open}), so the aggregate obligation defers to the runtime guard.

\begin{figure*}[t]
\centering
\includegraphics[width=0.62\textwidth]{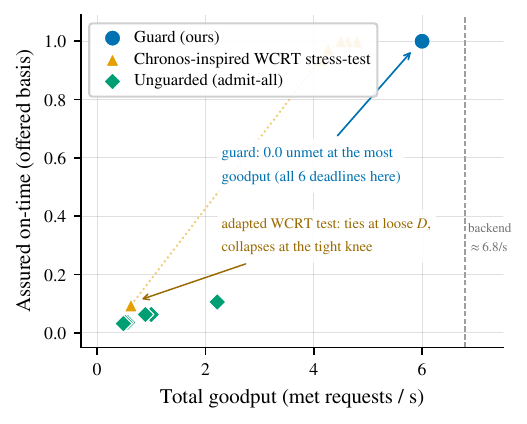}
\caption{\textbf{The guard is the only \emph{evaluated} arm at both the maximum assured on-time fraction \emph{and} maximum goodput under this grid} (offered basis, met\,/\,offered, so a rejected assured request counts against an arm),
at every deadline ($2\times$V100; $24$\,req/s offered, backend ${\approx}6.8$/s for this figure's mixed-size/deadline workload---distinct from \S\ref{sec:guard}'s ${\sim}16$-token saturation $\mu_{\mathrm{sat}}\!=\!9.6$--$9.7$/s---so the guard's $6$/s is
near-ceiling; reps $8$, the guard deterministic at $1.0$/$6$\,per\,s across reps, baseline per-rep spread small). The Chronos-inspired WCRT stress-test (our $C$-slot adaptation of Chronos Alg.~1, \emph{not}
Chronos itself) matches this fraction at loose deadlines but at \emph{lower} goodput and collapses at the tight
knee; admit-all loses goodput to self-congestion. Its reject-${\sim}$everything calibrations (${\approx}0$
goodput) are that adapted admitter's over-rejection regime, off this plane. \emph{Not} plotted, and not a weaker arm: a static \emph{per-class} cap ties the guard on a separate grid (same assured miss $0.0$ and opportunistic goodput): the per-class \emph{reservation} is the load-bearing ingredient, and a static \emph{count} cap gets it only \emph{at homogeneous sizes} (\S\ref{sec:guard}).}
\label{fig:pareto}
\end{figure*}

\FloatBarrier
\section{Related work}\label{sec:related}
\textbf{Emergence-as-Code}~\cite{krasnovsky2026emac} is closest in spirit: it compiles
a typed journey expression into optimistic/pessimistic availability and tail-latency
bounds and gates a rollout on the pessimistic bound. We share its meta-pattern and
claim no novelty there; we differ on \emph{object} (it composes per-service
availability along a journey and does \emph{not} model queueing/contention, which
are our subject) and on \emph{guarantee} (its bounds gate a \emph{pre-deployment}
rollout; we add a structural \emph{runtime} guard that protects a live request under an
online learner's mistakes). \textbf{Scheduling lineage (owned).} The guard's mechanism is classical: CBS-style
reservation~\cite{abeni1998cbs}, GPS/fixed priority~\cite{parekh1993gps}, and mixed
criticality's guaranteed floor under a misbehaving low-criticality load~\cite{vestal2007mixedcrit};
\S1's deterministic-vs-statistical split is the classical guaranteed-vs-statistical-QoS
distinction (network calculus~\cite{cruz1991calculus}). We take only runtime assurance's
\emph{stance}~\cite{sha2001simplicity,blackboxsimplex,neuralsimplex}, not its machinery; the
delta is the \emph{target} (an arbitrarily-wrong \emph{learned} admitter) and real hardware,
not a new scheduler. The \emph{enforcement} framing (edit
automata~\cite{ligatti2005edit}, Basin et al.'s controllable
partition~\cite{basin2013enforceable}) and the aggregate-tail
hyperproperty~\cite{clarkson2010hyperproperties} are cited for \emph{scope}, not novelty. \textbf{Bensalem et~al.}~\cite{threelayer2026} independently argue LLM-\emph{agent}
safety needs probabilistic assume-guarantee, but for semantic/operational predicates (they
scope out latency), via an information-availability seam, with no deterministic floor;
we complement it with the stochastic seam and an adversary-\emph{robust} floor (robust within
a characterized envelope, \S\ref{sec:guard}).
\textbf{Chronos}~\cite{chronos2026} is the nearest neighbor: it \emph{proves} a sound
WCRT admission test for single-GPU LLM serving ($0$ misses at $10\times$ load) and names
multi-node as its own open problem---but its schedulability model---a worst-case service-time model
(iteration-time coefficients $\alpha,\beta,\gamma$, Chronos's own, unrelated to our miscalibration $\beta$; + worst-case output length, \emph{not} a learned length predictor)---sits
in the \emph{trusted} path, and (like its $0$ misses) its guarantee is by-construction; our floor holds when
that model is instead \emph{arbitrarily wrong as a predictor} (we make no admission prediction at all) on
real $2\times$V100 (vs.\ its single-GPU simulation). The contrast is the \emph{predictor}, not the envelope:
our own bound still needs the \emph{actual} service to respect $s_{\max}$ (A3), an assumption about the
backend rather than about any model, and one we likewise do not claim as a proven WCET~(\S\ref{sec:open}).
The gap runs both ways: Chronos
proves a fuller \emph{token-level} (TTFT$+$TBT) WCRT theorem where we give a request-level
sojourn bound (\S\ref{sec:guard}); but adopting \emph{its} WCRT test as an admitter
(a conservative-admitter \emph{stress test}: a faithful re-implementation of its Alg.~1, $C$-slot-adapted, \emph{not} Chronos itself) over-rejects the
reserved floor, calibration-bound: a naive admission threshold $\tau$ over-rejects at the knee (miss $0.91$ at $600$\,ms) while a
sound $\tau$ admits $0$ for $D\!\le\!1200$\,ms (its $1/(1-\rho)$ busy-period bound blows up near $\rho\!=\!1$),
so no \emph{tested} $\tau$ achieved both in this grid, where our reservation admits \emph{unconditionally}. The contrast is goodput
$+$ calibration-robustness around an \emph{untrusted} admitter, not a protection horse-race; the
goodput--on-time contrast is Fig.~\ref{fig:pareto}.
A wrong learned \emph{router} (dispatch-class misclassifier) reduces to the same story: with $\rho\!<\!1$, even
dispatching by the \emph{mislabeled} class holds the floor ($0.0$ to a $0.5$ mislabel rate): dispatch order
bites only under overload. \textbf{SCORPIO}~\cite{scorpio2026} keeps a predictor in the trusted
path; our floor holds when the learner is arbitrarily wrong. Token pools~\cite{tokenpools2026}
and SLOs-Serve~\cite{slosserve2025} are single-layer token-accounting/allocation neighbors with
no untrusted-learner model (token pools \emph{itself} reserves for guaranteed workloads);
our delta is the cross-layer floor and that adversary model. \textbf{Kivi}~\cite{kivi2024} model-checks Kubernetes controllers \emph{and their configurations},
exhaustively exploring event interleavings to find violations; \textbf{Anvil}~\cite{anvil2024} verifies
controller \emph{code} (liveness). Neither reaches an ML-serving \emph{execution} model: our object is the
composition of that control-plane configuration with serving \emph{timing} (continuous batching, an untrusted
online admitter, a per-request deadline) and its SLO outcome. We claim no edge on the \emph{config} half: the
two hazards we mechanise (C4/C5, App.~\ref{app:hazards}) are pure quota arithmetic with no timing, and a
cluster-config model checker is the right tool for them. Gate-and-route~\cite{gateroute2026} proves fluid-limit
admission (predictor trusted), zero Kubernetes; the Semantic Router DSL~\cite{semrouter2026} verifies
routing well-formedness, not SLO. (We say ``floor,'' not ``envelope,'' to avoid
collision with~\cite{execenvelopes2026}.)

\section{An agenda: from a structural entitlement mechanism to a proved service guarantee}\label{sec:open}
\paragraph{Scope: what we can and cannot evaluate.}
As a Frontiers submission we are explicit about the evaluation's reach. We \emph{do} evaluate, on real
commodity hardware: the guard's floor under adversarially-miscalibrated learners (reps $10$, $2\times$V100);
the tested \emph{live} GAIE Flow Control configuration's sensitivity to an injected mapping fault (a real EPP v$1.5.0$ deployment); separately, inspection of that configuration found no authenticated, non-bypassable absolute reservation corresponding to our floor; and the static
screen's \emph{measured} false-accept regions over $1320$ hardware cells plus a held-out calibration. Three things are \emph{hard} to
evaluate and we do not claim them. \emph{(i)}~\textbf{Datacenter scale}: our testbed is a single commodity
$2\times$V100 node, so multi-node scale-out and datacenter accelerators are untested. \emph{(ii)}~\textbf{Real-model
Flow Control end-to-end}: we exercise FC's queueing mechanism on the official vLLM simulator (which isolates
exactly the priority-vs-reservation contrast at issue) rather than a production model server, and a head-to-head
with the guard on \emph{one} stack is future work. \emph{(iii)}~\textbf{A closed worst-case theorem}: continuous
batching couples service time to load (\S\ref{sec:guard}, A3), so the load-\emph{independence} a clean bound needs
does not hold in general; only the $B\!=\!C$ corner satisfies the conditional analytical bound (\S\ref{sec:guard}), and even there only by treating the calibrated $s_{\max}$ as a valid envelope, not a proven WCET. The demonstrated result is thus a
\emph{structure}; the four problems below turn it into a guarantee.

\paragraph{Discharge the capacity precondition.} SF holds only for backend capacity
$\ge\!B/W$ ($B$ credits per window, A4), and no guard conjures service. That precondition is itself a
composed-config fact (DRA placement, batch width, MoE expert availability), the cross-layer
property the static screen (\S\ref{sec:screen}) should monitor. A preliminary run bears this
out: a batch cap cutting throughput $62\%$ leaves the naive fixed floor at $3/120$ on time (full-class
\emph{unmet} $0.975$, all of it \emph{late}) where a lowered guard floor serves all it admits (unmet $0.833$,
all of it \emph{rejected}, none late): the two failures are not the same kind, which is why the shared
denominator is \emph{unmet} and not \emph{miss}; the screen \emph{accepted} the degraded floor \emph{before} the run, and the open mechanism makes it a guard that \emph{watches} its assumption online.

\paragraph{Close the worst-case guard theorem.} The \S\ref{sec:guard} sketch bounds assured
sojourn by a reserved-ahead term plus a non-preemptible tail $s^{\mathrm{opp}}_{\max}$; bounding
it worst-case (not worst-observed) needs a stochastic service model (continuous batching makes
service load-dependent). This theorem converts the empirical $0.0$ into a
proof; it is serving-specific and the paper's sharpest open problem. It is also
output-length-parametrized: the measured batch-$C$ service grows ${\approx}183+14.9\,\ell$ ms with output
$\ell$ (the clean same-batch median at $\ell\!\in\!\{64,128,256\}$; \S\ref{sec:screen}'s ${\approx}590$\,ms is
the batch-$C$ probe under \emph{sustained} load, ${\sim}170$\,ms higher as batching couples service to load),
so a $256$-token workload needs $D\!\gtrsim\!8$\,s: the cap bounds $s_{\max}$ at every length, but the floor's deadline grows with the output.

\paragraph{Make the screen sound near $\rho\!\to\!1$.} The $p99$ knob is an empirical
patch, not a proof: a \emph{sound} near-saturation screen needs a probabilistic model
checker or a proven headroom margin $\rho\le1-\epsilon$ trading admitted load for soundness.

\paragraph{Survive a moving upstream.} Whether trace-equivalence conformance holds as
the ecosystem absorbs a mechanism every one--two releases is an empirical question about
the upstream, not the theory, and that churn is live as of 2026-07: GAIE's EPP and
\texttt{InferenceObjective} have moved to the \texttt{llm-d/llm-d-router} project; the GAIE README labels the project \emph{GA} (its own roadmap still speaks of building ``towards a GA release''), and App.~\ref{app:hazards}'s C5 double-count is
still configuration-gated: current Kueue docs list its \texttt{KueueDRAIntegrationExtendedResource} mitigation as an \emph{alpha, default-off} gate (since v0.18), so unless an operator enables it the request and its auto-created claim can still be double-counted. Extending the guard to verified in-flight mode transitions
under DRA capacity loss is ongoing: a preliminary real-\texttt{DeviceTaintRule} run
load-shed gracefully, safety resting on a conservative degrade \emph{action}.

\section{Conclusion}
An assured SLO is not one obligation but two: a safety projection (its admission floor \emph{enforced} structurally, its service and drop properties conditional on stated service assumptions) that a
trusted guard upholds around a wrong learned admitter, and a
statistical residual a conservative screen can only \emph{approach}. Separating them lets a small trusted floor
bound an untrusted learner's damage: on real hardware the guard holds the assured class (\emph{admitted-basis}
miss \floornum{}) where an unguarded stack and a \emph{well-calibrated} FIFO admitter do not; a mechanism-level
test separately shows \emph{live} GAIE Flow Control dispatching by the priority our injected mapping fault sets (emulating an untrusted mapper; not a same-stack
head-to-head, nor an inherent GAIE limitation); and the cheap screen is honest about where only the
guard survives (the queueing knee, and for a $p50$ screen sub-saturation too). This is a \emph{Frontiers}
contribution: the stance and its early evidence are here; datacenter scale, real-model Flow Control, and a proved
worst-case floor are the agenda we invite the community to close.

\appendix
\section{The eight cross-layer hazards}\label{app:hazards}
Each row is a cross-layer surprise where every artifact is valid \emph{in isolation} but
the \emph{composed} QoS semantics surprise the operator, so \texttt{kubectl -{}-dry-run} and single-object OPA
(both single-object) catch \emph{none} of them. The rows differ in \emph{evidence level}, which the caption labels
per row: upstream issue reports (C1--C3, C6), a documented configuration semantics (C4) and a documented timing
gap (C7), a KEP mechanism whose double-count is our own derivation (C5), and our own batch-coupling measurement
(C8), so ``hazard'' spans reported bugs, documented behaviors, and risks we infer from composition, not a uniform
class of reproduced incidents. Our cross-layer quota checker mechanically detects C4/C5
(re-derived from the running \texttt{quota.py}, not hand-asserted); the rest are targeted by the transfer-function
approach but not yet mechanised (the honest split).

\begin{table*}[t]\centering\scriptsize\setlength{\tabcolsep}{4pt}
\begin{tabular}{@{}l l p{11.1cm} c@{}}
\toprule
Hazard & Layer pair & Composed cross-layer surprise (each artifact valid alone) & Checker \\
\midrule
C1 & Kueue cohort $\leftrightarrow$ kube \texttt{PriorityClass} & a user-reported cross-ClusterQueue preemption despite \texttt{reclaimWithinCohort:Never}; kube \texttt{PriorityClass} involved, root cause not definitively established & cat. \\
C2 & Kueue quota $\leftrightarrow$ preemption & \texttt{borrowWithinCohort:LowerPriority} lets a borrower preempt jobs running inside their own nominal quota & cat. \\
C3 & Kueue quota $\leftrightarrow$ GPU capacity & quota bookkeeping and actual multi-node \texttt{nvidia.com/gpu} availability diverge (``insufficient unused quota'' with capacity free) & cat. \\
C4 & Kueue quota-config $\leftrightarrow$ declared res. & \texttt{quotaCheckStrategy} silently skips a requested-but-undeclared resource $\Rightarrow$ a controller flag alone flips admission & \textbf{mech.} \\
C5 & DRA $\leftrightarrow$ Kueue quota & a device is charged twice (the \texttt{requests} alias \emph{and} its auto-created DRA claim) $\Rightarrow$ a job that physically fits is rejected & \textbf{mech.} \\
C6 & runtime timeout $\leftrightarrow$ service & a $10$\,s \texttt{default\_timeout} fires at ${\sim}90$\,s (wall-clock timeout $\ne$ dispatch-time feasibility; raw-$\mu$s knob) & cat. \\
C7 & Kueue admission $\leftrightarrow$ DRA binding & Kueue admits a workload \emph{before} the DRA device is bound (documented timing gap; \texttt{WaitForPodsReady} a documented safeguard) & cat. \\
C8 & runtime batching $\leftrightarrow$ admission sizing & continuous batching makes service load-dependent (${\sim}386$\,ms low-load vs ${\sim}460$\,ms under load, \S\ref{sec:guard}) $\Rightarrow$ a low-load-sized budget over-admits & cat. \\
\bottomrule
\end{tabular}
\caption{The eight cross-layer QoS hazards (\S\ref{sec:compose}). ``\textbf{mech.}'' $=$ mechanically
detected by the quota checker (verified vs the running \texttt{quota.py}); ``cat.'' $=$ catalogued, not yet
mechanised. Sources (evidence level; upstream status as of 2026-07): C1~Kueue~\#3210, C2~\#10171, C3~\#3909, C6~Triton~\#7512 (upstream issue reports); C4~\texttt{quotaCheckStrategy} (issue~\#7513 / PR~\#9808, a documented opt-in quota-check strategy, not a bug); C5~KEP-5004 (DRA extended-resource aliasing, beta in Kubernetes~v1.36; the double-count is our derivation of the failure the \texttt{KueueDRAIntegrationExtendedResource} gate prevents---current Kueue docs list that gate as \emph{alpha, default-off} since v0.18, so the mitigation is opt-in, not default); C7 a documented Kueue/DRA timing gap; C8 our own measurement (\S\ref{sec:guard}). \texttt{kubectl -{}-dry-run} and single-object OPA miss all eight.}
\label{tab:hazards}
\end{table*}

{\small \setlength{\bibsep}{1pt plus 0.2ex}\bibliographystyle{plainnat} \bibliography{refs}}
\end{document}